# Vulnerability Mitigation System (VMS): LLM Agent and Evaluation Framework for Autonomous Penetration Testing

Farzana Abdulzada

## Abstract
As the frequency of cyber threats increases, conventional penetration testing is failing to capture the entirety of todays complex environments. To solve this problem, we propose the Vulnerability Mitigation System (VMS), a novel agent based on a Large Language Model (LLM) capable of performing penetration testing without human intervention. The VMS has a two-part architecture for planning and a Summarizer, which enable it to generate commands and process feedback. To standardize testing, we designed two new Capture the Flag (CTF) benchmarks based on the PicoCTF and OverTheWire platforms with 200 challenges. These benchmarks allow us to evaluate how effectively the system functions. We performed a number of experiments using various LLMs while tuning the temperature and top-p parameters and found that GPT-4o performed best, sometimes even better than expected. The results indicate that LLMs can be effectively applied to many cybersecurity tasks; however, there are risks. To ensure safe operation, we used a containerized environment. Both the VMS and the benchmarks are publicly available, advancing the creation of secure, autonomous cybersecurity tools.

## KEYWORDS
**Autonomous Penetration Testing Agents, Benchmarks, Capture The Flag Challenges, Cybersecurity Automation, Large Language Models**

## INTRODUCTION
Penetration testing, a process of identifying and rectifying vulnerabilities by virtual attacks is a traditional practice that has been performed by human experts. However, as the systems are becoming more complex and the number of possible vulnerabilities to detect is increasing, this conventional approach has become too slow and resource-consuming. Although there are tools like Nessus, Snyk, and OpenVas that can detect vulnerabilities, they lack the flexibility required to address more complex scenarios.

In this paper, we present VMS - an autonomous agent powered by a Large Language Model (LLM) that can perform penetration testing without the need for human intervention. We also suggest new benchmarks for such agents. The purpose of this paper is to describe how LLMs can benefit cybersecurity tasks and how this area can be improved. This research addresses the problems of scalability and adaptability and highlights the importance of developing smart, safe, and effective tools to combat ever-emerging cyber threats and advance the creation of autonomous cybersecurity solutions.

## METHODS
This section describes the approach used to design and evaluate Vulnerability Mitigation System (VMS), an autonomous penetration testing agent trained by Large Language Models (LLMs) as well as the development of two standard CTF benchmarks. The study involves agent design, benchmark definition, and experimental analysis of LLM-based cybersecurity skills. The methodology consists of three main phases: the development of Vulnerability Mitigation System (VMS)'s architecture, the creation of the CTF benchmarks, and the performance testing and parameter optimization.

**Agent Architecture Design**
The Vulnerability Mitigation System (VMS) was developed with a dual-module architecture of a Planner and a Summarizer, both using Large Language Models for autonomous operation. The Planner generates executable commands from contextual data, using a system prompt that treats the LLM as an expert penetration tester, which solve CTF challenges. The Summarizer processes command outputs and maintains a history of actions to help guide subsequent decisions. Vulnerability Mitigation System (VMS) operates within a Kali Linux container, which is isolated by firewall rules to avoid any unintended network interactions. The workflow is an iterative cycle of the

Planner generating commands, executing them in the container and passing the output to the Summarizer, updating the action history, until a flag is captured or a step limit is reached. This phase was about creating a self-contained, independent problem solving system.

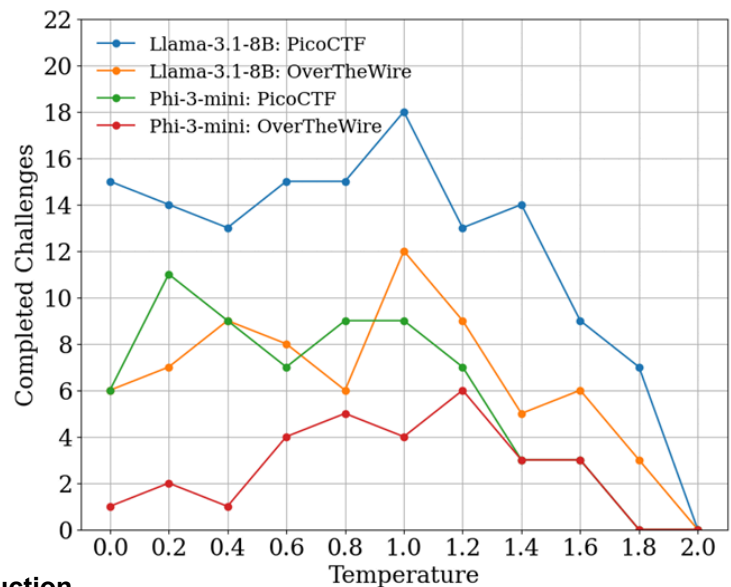

**Benchmark Construction**
Two sets of benchmarks for CTF were created using challenges from PicoCTF and OverTheWire, with a total of 200 tasks spread across six cybersecurity domains: General Skills, Cryptography, Web Exploitation, Forensics, Reverse Engineering and Binary Exploitation. From PicoCTF, 120 challenges were chosen, with descriptions, hints, and downloadable files, and the levels were classified as easy, medium, or hard. Solver scripts were written to pull flags dynamically. The benchmark included 80 challenges from four wargames — Bandit, Natas, Leviathan and Krypton — from OverTheWire, which helped to secure skills in Linux navigation, web security and cryptography. Both benchmarks were standardized to include metadata such as categories and difficulty levels, and solver functions were used to develop a reusable framework for assessing LLM-based agents.

**Experimental Evaluation**
The evaluation phase tested Vulnerability Mitigation System (VMS)'s performance across the benchmarks by optimizing three variables: observation window size (to limit the length of command output), temperature (to control response variability) and top-p (to control token selection diversity). Smaller LLMs such as Llama-3.1-8B and Phi-3-mini were also tested with window sizes of 250 (PicoCTF) and 500 (OverTheWire) characters to achieve a balance between context retention and clarity. Coherence was controlled by setting temperature to 1.0 and moderate creativity was ensured by fixing top-p to 0.9. GPT-4o, GPT-4o-mini, Llama-3.1-8B, Llama-3.1-70B, Qwen2-72B, Mixtral-8x7B, Phi-3-mini-4k and Phi-3.5-MoE were tested in the eight LLMs for 20 iterative steps per challenge. Metrics such as challenges solved, time per challenge and command usage patterns were collected and safety was assessed by monitoring for unintended behaviors using a containerized environment. The results were analyzed to compare model efficacy against domains and difficulty levels to gain insights into the suitability of LLMs for autonomous penetration testing.

## RESULTS AND DISCUSSION
Vulnerability Mitigation System (VMS)'s performance was evaluated across two benchmarks, with parameter optimization and benchmark runs yielding detailed insights. On the PicoCTF benchmark (120 challenges), GPT-4o solved 41 challenges, outperforming Llama-3.1-70B (27 challenges) and GPT-4o-mini (26 challenges), as shown in **Table 1.1**. On the OverTheWire benchmark (80 challenges), GPT-4o completed 32 challenges, followed by Llama-3.1-70B (23 challenges). Parameter optimization revealed an optimal observation window size of 250 for PicoCTF and 500 for OverTheWire, with performance peaking at a temperature of 1 and top-p of 0.9 (**Figure 1.1**). Higher temperatures increased error rates, with commands at 2.0 rendering systems unusable. Token usage rose linearly with steps, plateauing for some models after 10 iterations.

**Table 1.1** Performance of LLMs on PicoCTF and OverTheWire Benchmarks

| LLM | PicoCTF (120) | OverTheWire (80) | Avg. Time/Challenge (s) |
|---|---|---|---|
| **GPT-4o** | 41 | 32 | 15.2 |
| **Llama-3.1-70B** | 27 | 23 | 18.7 |
| **GPT-4o-mini** | 26 | 19 | 16.8 |

## CONCLUSIONS

In this paper, we propose the Vulnerability Mitigation System (VMS), a completely autonomous penetration testing agent, which is backed by a Large Language Model (LLM). The system was able to solve a vast majority of the Capture The Flag (CTF) challenges without any form of human intervention. The VMS was designed to have two modules; the Planner and the Summarizer for executing commands in a step by step manner. Among the models evaluated, GPT-4o was the most effective, it was able to solve 41 of the 120 PicoCTF challenges and 32 of the 80 OverTheWire challenges, which was beyond our expectations. We also released two new benchmarks consisting of 200 challenges in total to further assess the system and determined the optimal settings such as using temperature 1 and observation window sizes of 250 and 500 respectively.

The results also show how LLMs can enhance cybersecurity automation by addressing the problems of scalability and adaptability. But we noticed some safety risks, for example, the possibility of inappropriate use of the system at higher temperatures, therefore we need strong safety measures like containerization. This research therefore becomes the first to propose the VMS and its benchmarks for public access in order to enable future improvement of autonomous cybersecurity tools.